\documentclass[aps,prl,twocolumn,preprintnumbers,groupedaddress,superscriptaddress,floatfix,tightenlines,reprint]{revtex4-1}
\usepackage{mathrsfs,natbib}
\usepackage{graphics,epsfig,color,graphicx}
\usepackage{verbatim}
\usepackage{bm}
\usepackage{enumerate}

\usepackage{graphicx,epsf,amssymb,bbm,amsbsy,amsfonts,amssymb,amsmath}
\usepackage{hyperref}

\newcommand{\eq}{\begin{equation}}
\newcommand{\eqe}{\end{equation}}

\def\tr{\text{tr}\,}

\def\Re{\mathrm{Re}\,}

\hypersetup{
    colorlinks=true,       
    linkcolor=red,          
    citecolor=blue,        
    filecolor=magenta,      
    urlcolor=blue           
}

\input xy
\xyoption{all}

\newcommand{\bea}{\begin{eqnarray}}
\newcommand{\eea}{\end{eqnarray}}

\newcommand{\oneover}[1]{\ensuremath{\frac{1}{#1}}}
\newcommand{\half}{\oneover{2}}

\newcommand{\nf}{{n_{\rm f}}}
\newcommand{\U}{{\mathcal{U}}}
\newcommand{\omegaf}{\nu}

\begin{document}

\title{Order parameters and color-flavor center symmetry in QCD}

\preprint{INT-PUB-17-021}

\author{Aleksey Cherman}
\email{aleksey.cherman.physics@gmail.com}
\affiliation{Institute for Nuclear Theory, University of Washington, Seattle, WA 98105 USA}
\author{Srimoyee Sen}
\email{srimoyee08@gmail.com}
\affiliation{Department of Physics, University of Arizona, Tucson, AZ 85721 USA}
\author{Mithat \"Unsal}
\email{munsal@ncsu.edu}
\affiliation{Department of Physics, North Carolina State University, Raleigh, NC 27695 USA}
\author{Michael L. Wagman}
\email{mlwagman@uw.edu}
\affiliation{Institute for Nuclear Theory, University of Washington, Seattle, WA 98105 USA}
\affiliation{Department of Physics, University of Washington, Seattle, WA 98105 USA}
\author{Laurence G. Yaffe}
\email{yaffe@phys.washington.edu}
\affiliation{Department of Physics, University of Washington, Seattle, WA 98105 USA}

\begin{abstract}
Common lore suggests that $N$-color QCD with massive quarks
has no useful order parameters which can be non-trivial
at zero baryon density.
However, such order parameters do exist when
there are $\nf$ quark flavors with a common mass
and $d\equiv\gcd(\nf,N) > 1$.
These theories have a $\mathbb Z_d$ 
color-flavor center symmetry arising from intertwined
color center transformations and cyclic flavor permutations.
The symmetry realization depends on
the temperature, baryon chemical potential and value of $\nf/N$,
with implications for 
conformal window studies and dense quark matter.
\end{abstract}
\maketitle

{\bf Introduction.} 
Defining order parameters in QCD is notoriously subtle.   
In pure $SU(N)$ Yang-Mills (YM) theory,
the simplest non-trivial order parameter is the expectation value
of a line operator:
\begin{align}
\langle \tr \Omega \rangle = \langle\tr \, \mathcal{P} \, e^{i\int_{0}^{L} dx_1 A_1} \rangle \,,
\label{eq:Polyakov_loop}
\end{align}
when the $x_1$ dimension is compactified with circumference $L$.
If $x_1$ is regarded as Euclidean time,
then the gauge theory functional integral with periodic boundary conditions
calculates the thermal partition function
with temperature $T = 1/L$.
The thermal expectation value \eqref{eq:Polyakov_loop} is the Polyakov
loop confinement order parameter for the $\mathbb{Z}_{N}$ center symmetry
of pure YM, whose realization changes with temperature.

Adding fundamental representation quarks $\{q_a\}$, $a=1,2,{\cdots}, \nf$ explicitly breaks the $\mathbb Z_N$ center symmetry and complicates the story.  The  Polyakov loop ceases to be an order parameter.  With massless quarks, the flavor symmetry is $G = SU(\nf)_V \times SU(\nf)_A \times U(1)_Q$ and the chiral condensate $\langle \sum_a \bar{q}_a q_a \rangle$
is an order parameter for the $SU(\nf)_A$ chiral symmetry,
whose realization depends on temperature.
But if the quarks are massive, as in real-world QCD,
then chiral symmetry is explicitly broken and
$\langle \bar{q}_a q_a \rangle$ ceases to be an order parameter.
The remaining vector-like $U(\nf)$ symmetry cannot break
spontaneously at vanishing baryon number density \cite{Vafa:1983tf},
leading to the common understanding that
QCD with dynamical massive quarks lacks non-trivial order parameters at zero baryon density.

This standard lore overlooks the possibility
of
 symmetries which intertwine center and flavor transformations.
Our discussion generalizes earlier work 
\cite{Kouno:2012zz,
    Sakai:2012ika,
    Kouno:2013zr,
    Kouno:2013mma,
    Iritani:2015ara,
    Kouno:2015sja,
    Hirakida:2016rqd,
    Hirakida:2017bye},
showing that 
special boundary conditions (BCs) for quarks
can lead to an unbroken $\mathbb Z_3$ symmetry.
These works interpreted this choice as defining a ``QCD-like" theory
that they termed ``$Z_3$-QCD".   
Other works have considered different applications of the same BCs~\cite{Briceno:2013hya,Cherman:2016hcd,Liu:2016yij,Cherman:2016vpt}.
Here, we generalize and reinterpret the constructions of these BCs
and use them to define
order parameters for both quantum and thermal phase transitions in QCD.

{\bf Color-flavor center symmetry.}
Center symmetry \cite{Weiss:1980rj}
acts only on topologically
non-trivial observables. 
It can be viewed as a topologically non-trivial gauge transformation, with the action 
\begin{equation}
    \langle \tr \Omega \rangle \to \omega \, \langle \tr \Omega \rangle
    \,,\qquad
    \omega \equiv e^{2\pi i/N} \,.
\end{equation}
By itself, this is not a symmetry when
fundamental representation fields are present in the theory.

We assume $\nf$ quark flavors have a common mass $m_q$,
so the theory (on $\mathbb R^4$) has a $U(\nf)_V$ flavor symmetry.
We consider flavor-twisted quark boundary conditions,
\begin{align}
    q_{a}(x_1{+}L) = \mathcal{U}^{ab} \, q_{b}(x_1) \, ,
\label{eq:generic_twisted_BCs}
 \end{align} 
where $\mathcal{U}$ is a $U(\nf)$ matrix,
and regard $x_1$ as a \emph{spatial} direction. 
The flavor twist $\mathcal U$ may be assumed diagonal without
loss of generality.
The $SU(\nf)_V$ flavor subgroup has center $\mathbb{Z}_{\nf}$,
which motivates a $\mathbb Z_{\nf}$ symmetric choice of BCs
\cite{Kouno:2012zz,
    Sakai:2012ika,
    Kouno:2013zr,
    Kouno:2013mma,
    Iritani:2015ara,
    Kouno:2015sja,
    Hirakida:2016rqd,
    Hirakida:2017bye}
for which the set of eigenvalues of $\mathcal U$ is invariant under
multiplication by elements of $\mathbb Z_{\nf}$.
If the theory is to retain charge conjugation
and $x_1$-reflection symmetries (suitably redefined),
then the set of eigenvalues must also
be invariant under complex conjugation.
Two possibilities result, namely $\nf$'th roots of $+1$ or $-1$,%
\begin{subequations}\label{eq:BCs}%
\begin{align}
    \mathcal{U} &= \textrm{diag}(1, \omegaf, \cdots, \omegaf^{\nf-1}),\qquad
    \omegaf \equiv e^{2\pi i/\nf} \,,
\label{eq:BCa}
\\
\noalign{\hbox{or}}
    \mathcal{U} &= \textrm{diag}(\omegaf^{1/2}, \omegaf^{3/2}, \cdots,
    \omegaf^{\nf-1/2}) \,.
\label{eq:BCb}
\end{align}
\end{subequations}
With the BCs in \eqref{eq:BCs}, the finite $L$ flavor symmetry is reduced to
$G_{\rm L} = U(1)^{\nf-1}_V \times U(1)^{\nf-1}_A \times U(1)_Q \subset G$.

The key observation is that if
\begin{equation}
    d \equiv \mathrm{gcd}(\nf, N) > 1 \,,
\end{equation}
then the circle-compactified theory, with either boundary condition
(\ref{eq:BCs}),
also remains invariant under an intertwined $\mathbb Z_d \subset \mathbb{Z}_{N} \times \mathbb{Z}^{\rm perm}_{\nf}$ color-flavor center (CFC)
symmetry, generated by the combination of a center transformation
with phase $\omega^{N/d} = e^{2\pi i/d}$ and a $\mathbb Z_d$ cyclic
flavor permutation.
To see this note that, given either choice (\ref{eq:BCs}),
a $\mathbb{Z}_d$ center transformation effectively permutes the
eigenvalues of $\mathcal U$.
Combining the center transformation with the opposite cyclic flavor
permutation (which is part of the $U(\nf)_V$ flavor symmetry) leaves the boundary condition invariant.

CFC symmetry intertwines center and flavor transformations and so has both local and extended order parameters.
Examples of CFC order parameters include 
 Polyakov loops such as \eqref{eq:Polyakov_loop}
with winding numbers which are non-zero mod $d$,
as well as $\mathbb Z_\nf$ Fourier transforms of fermion bilinears,
$
    \mathcal{O}^{(p)}_{\Gamma}
    \equiv
    \sum_{a=1}^{\nf} \omegaf^{-a p} \,
    \bar{q}_{a} \Gamma q_{a}
$,
where $\Gamma$ is an arbitrary Dirac matrix and $p \bmod d \ne 0$. The action of the $\mathbb Z_d$ CFC symmetry is
\begin{align}
\label{eq:CFC_action}
\begin{split}
    \tr \Omega^{p} \;&\to \; \omega^{N p/d} \>
    \tr \Omega^{p}\, ,
    \quad
    \mathcal{O}^{(p)}_{\Gamma} \;\to \; \nu^{\nf p/d} \>
    \mathcal{O}^{(p)}_{\Gamma} \,.
\end{split}
\end{align}
Other related choices of boundary conditions,
and generalizations to multiple compactified directions
are discussed in our Supplemental Materials.

{\bf Center symmetry and confinement.}
Consider the Polyakov
loop connected correlator in QCD compactified  on $x_1$ with circumference $L$,
\begin{align}
    \langle \tr \Omega(\vec{x}) \, \tr \Omega^{\dag}(0) \rangle_{\rm conn}
    \equiv e^{-F(r)}
    \,, \qquad r = |\vec{x}| \,.
    \label{eq:polyakov_correlator}
\end{align} 
Suppose there is a non-zero lower bound $E$ on the energy  of states that can contribute to the correlator, so $F(r) \sim E\, r$ as $r \to \infty$.
When $\nf = 0$, the theory has a $\mathbb{Z}_N$ center symmetry.  If the ground state is $\mathbb{Z}_N$ invariant, then no intermediate state created by a local operator acting on the vacuum
can contribute to the correlator.
All contributions to the correlator
\eqref{eq:polyakov_correlator} must involve flux tubes which wrap
the compactified dimension, so that $E = L \sigma$
with $\sigma$ the string tension.   

On the other hand, if center symmetry is broken,
explicitly or spontaneously,
then intermediate states 
created by local operators can also contribute to the correlator
\eqref{eq:polyakov_correlator}.
The minimal energy $E$ need not grow with  $L$.
This is interpreted as a signal of string breaking.
It is tempting to conclude that there is a tight link between
unbroken center symmetry and confinement of static test quarks
by unbreakable flux tubes.  

Now suppose that $d=\mathrm{gcd}(\nf, N) >1$, all quarks have a common mass $m_q$,
and we engineer the existence of $\mathbb{Z}_d$ CFC symmetry by using the BCs \eqref{eq:BCs}.
As seen above,
CFC symmetry acts on both Polyakov loops and appropriate local operators.
Intermediate states created by
local operators transforming the same as $\tr\Omega$ under all unbroken
symmetries can contribute to the correlator
\eqref{eq:polyakov_correlator}.
For example,
states created by
$
    \mathcal{A} \equiv
    \sum_{a=1}^{\nf} \omegaf^{-a p} \,\bar{q}_{a} \gamma_{1} D_1 q_{a}
$
and
$
    \mathcal{B} \equiv
    \sum_{a=1}^{\nf} \omegaf^{-a p} \, \bar{q}_{a} \gamma_{1} q_{a}
$
can contribute to correlators of $\mathrm{Re} \, \tr \Omega^p$
and $\mathrm{Im} \, \tr \Omega^p$, respectively,
even when CFC symmetry is not spontaneously broken.
Consequently, the string tension as defined by the asymptotic behavior of the correlator
\eqref{eq:polyakov_correlator}
vanishes regardless of the realization of the $\mathbb{Z}_d$ center symmetry.
Of course, the minimal masses of mesons created by operators $\mathcal A$
or $\mathcal B$ grow with increasing quark mass $m_q$.
Due to non-uniformity in the 
$m_q \to \infty$ and $r \to \infty$ limits, a non-zero string
tension does emerge if one sends $m_q \to \infty$ first.
In summary, we  see that for $\nf>0$ there is no relation between
the presence of a non-zero string tension and
the existence, or realization, of a center symmetry intertwined with flavor.

{\bf Conformal window.}  
\begin{figure}
\centering
\includegraphics[width=0.35\textwidth]{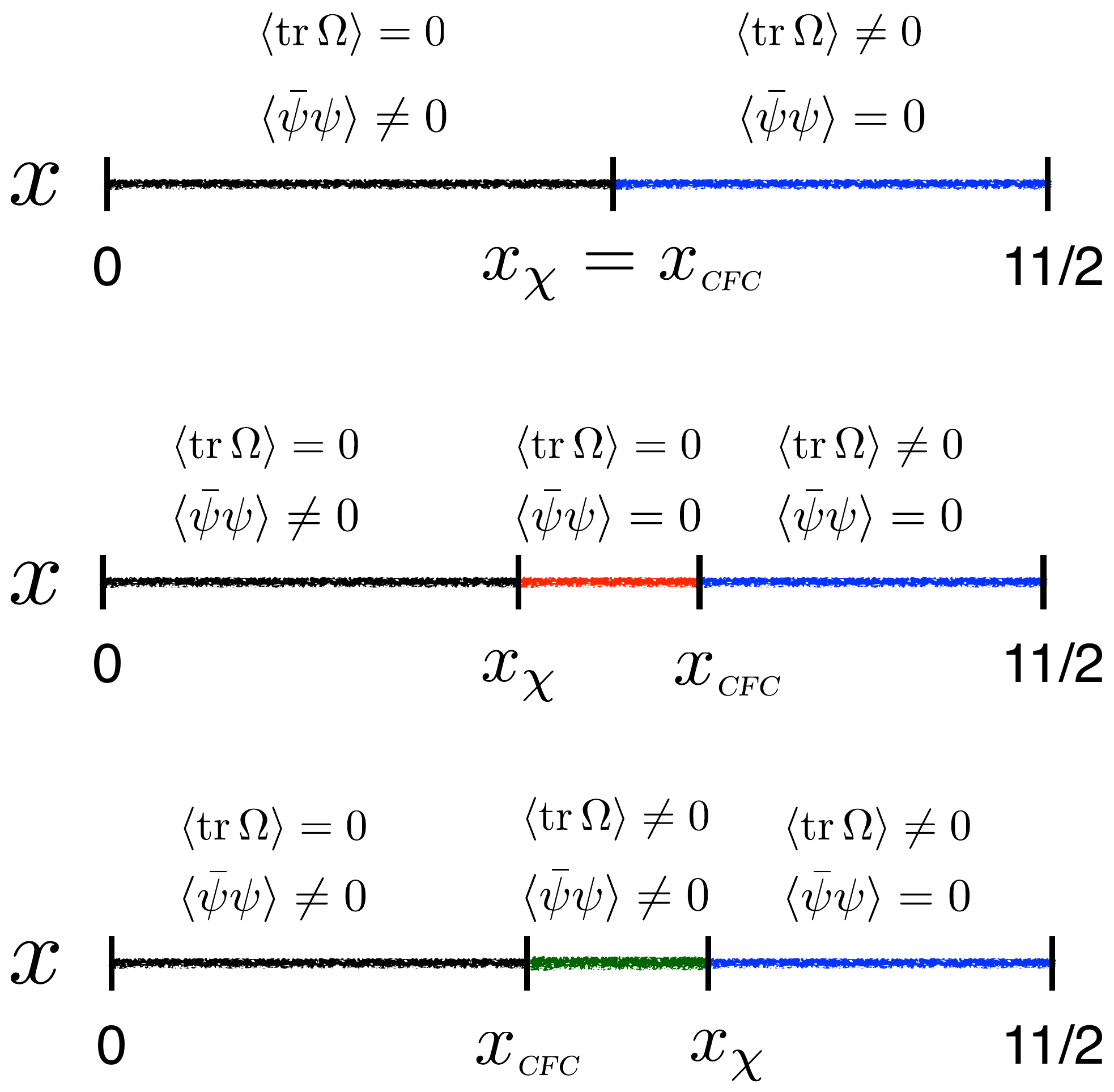}
\caption{Possible phase structures of massless QCD as a
function of $x = \nf/N$.
The chiral and CFC symmetry realizations change at some $x = x_{\chi}$ and $x = x_{\rm CFC}$, respectively.
}
\label{fig:conformal_window}
\end{figure}
Let  $x \equiv \nf/ N$, and set to zero the common quark mass and
temperature, $m_q=T=0$.
If $x>\frac{11}{2}$, QCD  becomes an infrared-free theory.
For $x$ below some $x_{\chi} < \frac{11}{2}$, chiral symmetry is believed
to be spontaneously broken.
In the intermediate range of values
$x \in (x_{\chi}, \frac{11}{2})$,
called the conformal window,
QCD flows to a
non-trivial infrared (IR) fixed point without chiral symmetry breaking.
The value of $x_{\chi}$ has been the subject of intensive
lattice investigations
(see, e.g., Refs.~\cite{Aoki:2014oha,Rinaldi:2015axa,Appelquist:2016viq,Fodor:2009wk,Aoki:2013xza,Appelquist:2014zsa,Appelquist:2012nz,Fodor:2011tu,Aoki:2012eq,Aoki:2013zsa}).
The existence of an IR-conformal phase can be seen most easily in
the Veneziano large $N$ limit of QCD, where $x$ is fixed along with
the 't Hooft coupling $\lambda \equiv g^2 N$ as  $N$ increases.
If $\epsilon \equiv \frac{11}{2}-x \to 0^{+}$, perturbation
theory self-consistently implies the existence of an IR fixed
point with a parametrically small coupling
\cite{Caswell:1974gg,Banks:1981nn}, $\lambda_{\rm IR} = \tfrac{64}{75}
\pi ^2 \epsilon\ll 1$.

One may show
that $\mathbb Z_d$ CFC symmetry is
spontaneously broken in the conformal window,
at least at large $N$.
The Veneziano limit is taken through a sequence of values
for which $d = \gcd(\nf,N)$ is fixed and greater than 1,
while the ratio $x=\nf/N$ approaches a non-zero limit.
Hence, the Polyakov loop \eqref{eq:Polyakov_loop} remains
an order parameter for the intertwined $\mathbb{Z}_d$ center symmetry.

The CFC realization may be determined by computing the
quantum effective potential $V_{\rm eff}(\Omega)$.
The loop expansion is 
controlled by the small value of $\lambda$ at all scales when
$\epsilon \ll 1$, rather
than the small size of $L$ compared to the inverse strong scale $\Lambda^{-1}$
as in the classic papers \cite{Gross:1980br,Weiss:1980rj}.
Hence, the following analysis is valid for any circumference $L$,
including the $L \to \infty$ limit of interest.

Classically, $V_{\rm eff}(\Omega)$ is zero.
Using standard methods
\cite{Gross:1980br,Weiss:1980rj},
at one loop one finds
$V_{\rm eff}(\Omega) = V_{\rm g}(\Omega) + V_{\rm f}(\Omega)$
with gluon and fermion contributions given by
\begin{align}
\label{eq:Vg}
    V_{\rm g}(\Omega) &=
    -\frac{2}{\pi^2 L^4}  \sum_{n=1}^{\infty} \frac{1}{n^4}
    \left(|\tr \Omega^n|^2-1\right) ,
\\
\noalign{\hbox{and}}
\label{eq:Vf}
    V_{\rm f}(\Omega) &=
    \frac{2}{\pi^2 L^4}\sum_{n=1}^{\infty} \frac{1}{n^4}
    \left(  \tr \U^{-n} \>\tr \Omega^n +  \tr \U^{n} \>\tr \Omega^{-n} \right)
\nonumber
\\
    &= \frac{2}{\pi^2 L^4 \nf^3}\sum_{n=1}^{\infty} \frac{(\pm1)^n}{ n^4}
    \left(\tr \Omega^{\nf n} +\mathrm{h. c.} \right) \, .
\end{align}
The upper/lower sign refers to BCs \eqref{eq:BCa}/\eqref{eq:BCb}.
As required, $V_{\rm eff}$ is invariant under CFC symmetry.
To determine the minima of $V_{\rm eff}$
note that $V_{\rm g} = O(N^2)$ while,
due to our imposition of flavor-twisted BCs, $V_{\rm f} = O(N^{-2})$.
At large $N$,
the minima of $V_{\rm eff}$ are entirely determined by the gluonic
contribution $V_{\rm g}$, which favors coinciding eigenvalues,
$\Omega\propto 1$.
Consequently, when $\epsilon = \frac {11}{2}{-}x \ll 1$ 
the CFC symmetry is spontaneously broken at any $L$.
On the other hand,
at the pure Yang-Mill point, $x=0$,
center symmetry is certainly expected to be unbroken
at large $L$, and standard
large $N$ counting arguments imply that the intertwined center symmetry
should remain unbroken for sufficiently small $x$.
Hence, there must be at least one transition at some $x = x_{\rm CFC}$
where the realization of the CFC symmetry changes.
This point may or may not coincide with the point $x_{\chi}$
where the chiral symmetry realization changes.
Logically possible phase diagrams are sketched in
Fig.~\ref{fig:conformal_window}.

Introducing a non-zero quark mass or temperature gives a richer
phase structure.
With $\epsilon \ll 1$ and small quark mass,
the theory develops a new strong scale,
$\Lambda_{m} \sim m_{q} \, e^{-75/(8 \epsilon^2)}$.
As $L$ increases and becomes comparable to $\Lambda_m^{-1}$
we expect a $\mathbb Z_d$ center-restoring phase transition.
We also expect a CFC-restoring phase transition at a non-zero temperature
$T_{*} \sim 1/L$ when $m_q=0$,
similar to the large $N$ deconfinement transition in
$\mathcal N\,{=}\,4$ super-Yang-Mills theory on $S^3\times S^1$
\cite{Witten:1998qj,Aharony:2003sx}.

Now consider $N\,{=}\,3$ and massless quarks.
If the $\nf = 15$ IR fixed point is weakly coupled,
as widely believed, then our above calculation applies
and $\mathbb{Z}_3$ center symmetry is spontaneously broken at $\nf = 15$.
At $\nf \,{=}\, 3$, lattice calculations
\cite{Iritani:2015ara} with boundary conditions \eqref{eq:BCa}
are consistent with unbroken $\mathbb Z_3$ center symmetry when
$L \Lambda \gg 1$.
So for integer values of $x = \nf/3$, there must be a minimal value
$2 \le x_{\rm CFC} \le 5$ where the $\mathbb Z_3$ CFC symmetry
first becomes spontaneously broken.

{\bf Dense quark matter.}
Consider the phase diagram of QCD
with $N = \nf = 3$ and a common quark mass $m_q$, as a function of the $U(1)_Q$ 
chemical potential $\mu$ and temperature $T$.
Previously known symmetry principles only suggest the existence of a curve
$T(\mu)$ in the $(T,\mu)$ plane below which lies a superfluid phase with
spontaneously broken $U(1)_Q$ symmetry,
leading to a hypothesis of continuity of quark matter and hadronic nuclear
matter \cite{Schafer:1998ef}.
Consideration of CFC symmetry implies the existence of additional phase structure when QCD  is compactified with CFC-preserving BCs on a spatial circle large compared to other spatial scales.

\begin{figure}
\centering
\includegraphics[width=0.35\textwidth]{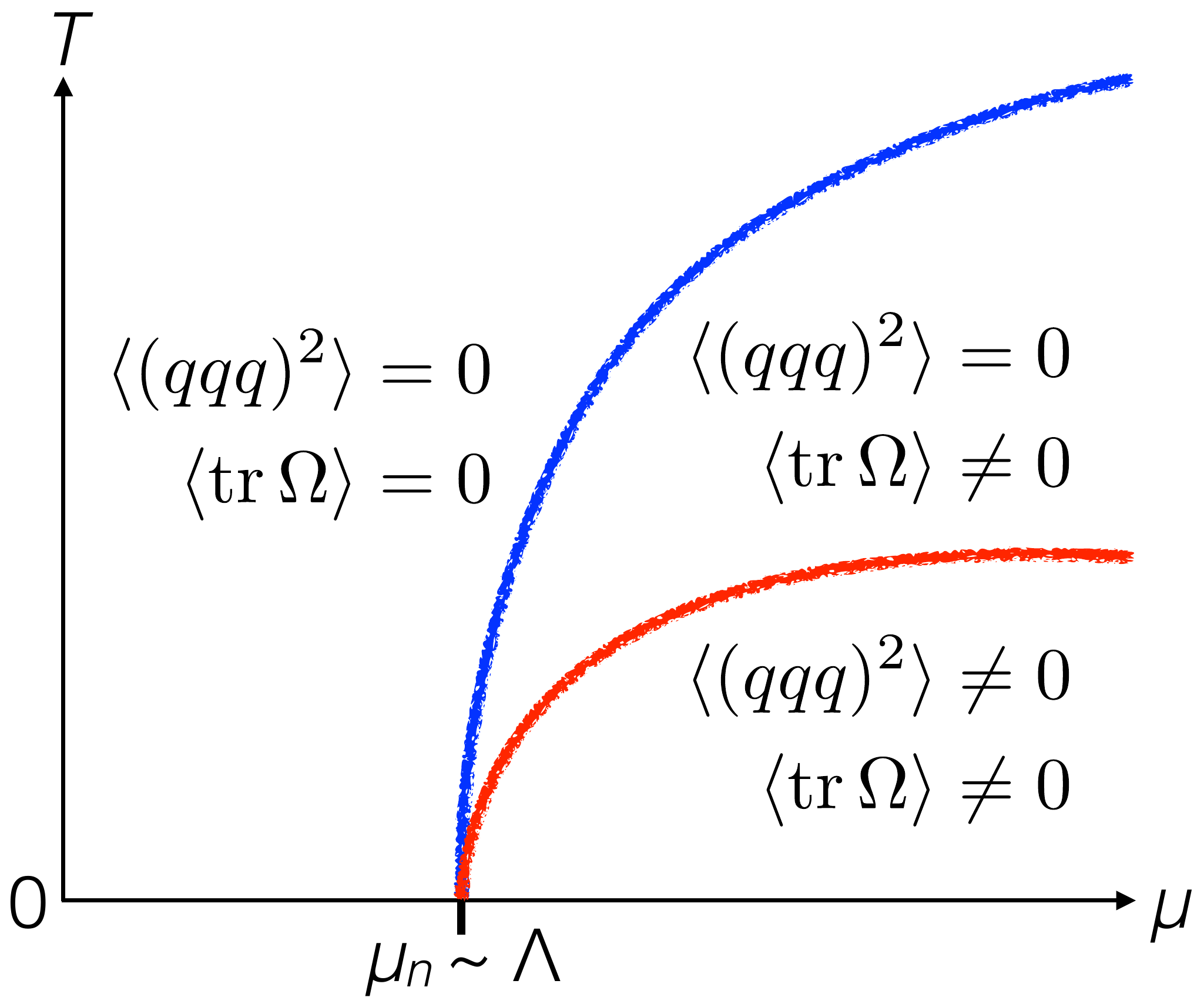}
\caption
    {(Color online.)
    Sketch of a possible phase diagram of circle-compactified
    $SU(3)_V$ symmetric QCD at $m_q>0$,
    as a function of $T$ and $\mu$, in the large $L$ limit.
    }
\label{fig:phaseDiagram}
\end{figure}

First, consider the small $T$, small $\mu$ regime.  Here
lattice studies \cite{Iritani:2015ara} imply that
$\langle \tr \Omega \rangle = 0$ at large $L$.
Next, consider high temperatures, $T\gg \mathrm{max}(\Lambda, \mu)$.
Here, the dynamics on spatial scales large compared to $(g^2 T)^{-1}$ are
described by pure 3D YM theory \cite{Gross:1980br}
which confines, so $\langle \tr \Omega \rangle \,{=}\, 0$ at high $T$.
We expect this high-temperature region to be smoothly connected to the region near $T = \mu=0$.
However, as we next discuss the CFC symmetry realization behaves non-trivially
when $T \to 0$ with $\mu \gg \mathrm{max}(\Lambda, m)$.
A simple phase diagram consistent with our results is sketched in 
Fig.~\ref{fig:phaseDiagram}.

High density QCD, $\mu \gg \Lambda$,
is believed to be in a ``color-flavor-locked" (CFL)
color-superconducting phase \cite{Alford:2007xm}
when $T <T_{\rm CFL}$.
The phase transition temperature $T_{\rm CFL}$ is comparable
to the superconducting gap,
$T_{\rm CFL} \sim \Delta \sim \mu \, g^{-5} e^{-(3\pi^2/\sqrt 2)/g}$.
Electric and magnetic gluons develop Debye and Meissner static screening
masses, respectively, both of order $g\mu$
in the CFL phase \cite{Son:1999cm,Rischke:2000ra}.
For $T_{\rm CFL} < T \lesssim g\mu$, low frequency magnetic fluctuations
experience Landau damping.
Consequently, for $T \lesssim g\mu$
the relevant gauge coupling is small,
$g(\mu) \ll 1$,
and cold dense quark matter is weakly coupled.

\begin{figure}
\centering
\includegraphics[width=.5\textwidth]{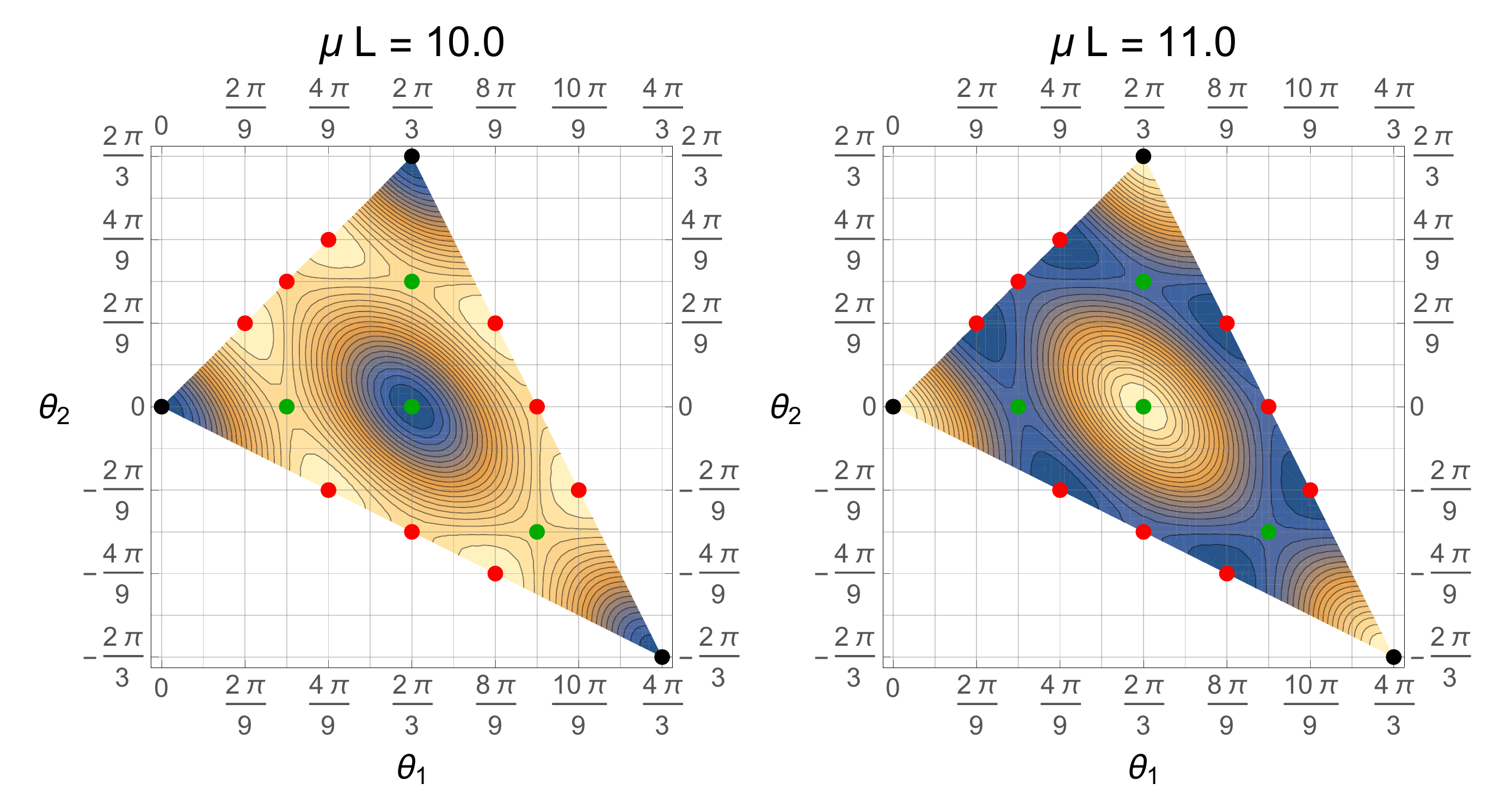}
\vspace*{-10pt}
\caption
    {(Color online.)
    Contour plots of $V_{\rm f}$ for $N = \nf = 3$,
    with BCs \eqref{eq:BCa},
    as a function of $\theta_1,\theta_2$ for two nearby values
    of $\mu L$ with $T/\mu = 10^{-3}$, illustrating the quantum oscillations described in the text.
    Darker colors indicate lower values of $L^4 V_{\rm eff}$.
    Regions outside the triangle shown are gauge-equivalent to points within
    the triangle.
    The center-symmetric point
    $(\theta_1,\theta_2,\theta_3)=(0,2\pi/3,4\pi/3)$ lies at the center
    of the triangle while the corners are the coinciding eigenvalue
    points $(0,0,0)$ and $\pm(2\pi/3,2\pi/3,2\pi/3)$.
    Dots denote critical points of $\widehat{V}_{\rm f}$.
    Results with BCs \eqref{eq:BCb} are similar.
    }
\label{fig:quantum_oscillation}
\end{figure}

In typical gauge-dependent language, CFL superconductivity is driven
by an expectation value for diquark operators,
$
    \langle q^{a}_{i} C \gamma_5 q^{b}_{j} \rangle
    \propto
    \epsilon^{abK} \epsilon_{ijK}
$
\cite{Alford:2007xm}.
The uncontracted flavor indices on the ``condensate" might lead
one to think that flavor permutation symmetry is broken,
automatically implying accompanying spontaneous
breaking of the CFC symmetry \cite{Kouno:2015sja}
when $x_1$ is compactified with BCs \eqref{eq:BCs}.
But this gauge-dependent language is misleading.
The true gauge-invariant order parameters for
spontaneous breaking of chiral and $U(1)_Q$ symmetries,
schematically
$\langle \bar{q} C \gamma_5 \bar{q} q C \gamma_5 q \rangle$
and $\langle (q C \gamma_5 q)^3 \rangle$,
are $SU(\nf)_V$ singlets \cite{Alford:2007xm}.
So the development of CFL superconductivity does not, ipso facto,
imply spontaneous breaking of CFC symmetry in the
circle compactified theory.

To study the CFC symmetry realization 
when $\mu \gg \Lambda$ and $T\ll g\mu$,
we examine the one-loop effective potential for $\Omega$,
which is the sum of gluonic and fermionic contributions,
$
    V_{\rm eff}(\Omega)
    =
    V_{\rm g}(\Omega) + V_{\rm f}(\Omega)
$.
The loop expansion is controlled by $g(\mu)\ll 1$ and is applicable for all $L$.
For $T < T_{\rm CFL}$,
gluons have effective masses $m_{\rm g} \sim g \mu$ due to a combination
of Debye screening and the Meissner effect from color superconductivity
\cite{Son:1999cm,Rischke:2000ra}.
As with any one-loop holonomy effective potential contribution from massive adjoint bosons, we thus expect
\begin{align}
    V_{\rm g}(\Omega) =
    - \frac 1{L^{4}} \sum_{n=1}^{\infty} \> f_n \, (|\tr \Omega^n|^2-1) \,,
\label{eq:Vg2}
\end{align}
with coefficients $f_n > 0$ which are exponentially small,
$f_n \sim e^{-n m_{\rm g} L}$, when $m_{\rm g} L \gg 1$.
At large $\mu$, $V_{\rm g}(\Omega)$
is highly suppressed compared to
the $\mu=0$ result \eqref{eq:Vg}.

Fermion excitations near the Fermi surface are nearly gapless as $T \to 0$,
up to non-perturbative corrections from quark pairing.
Interactions are weak, so we are dealing with a nearly free Fermi liquid. 
In the cold dense limit, $V_{\rm eff}(\Omega)$ is completely
dominated by the fermion contribution,
\begin{align}
\label{eq:Vfgeneral}
    V_{\rm f}(\Omega)
    &= \frac{1}{\pi \beta L^3 \nf^2}
    \sum_{n=1}^\infty
      \frac{(\pm 1)^n}{n^3} \,
       \Bigl[ \left(\tr \Omega^{\nf n}  +\mathrm{h.c.} \right) 
\nonumber
\\ & \qquad\qquad {} \times
    \sum_{k\in \mathbb{Z}+\half}
      (1+\nf n \mathfrak{m}_k L) \>
      e^{-\nf n  \mathfrak{m}_k L}
      \Bigr] \,,
\end{align}
where
$
    \mathfrak{m}^2_k 
    \equiv
    (2\pi k T + i\mu)^2 + m_q^2
$,
and the upper/lower sign refers to BCs \eqref{eq:BCa}/\eqref{eq:BCb}.
(Derivation detailed in Supplemental Materials.)
This result is manifestly invariant under the CFC symmetry, as required.
To examine the realization of CFC symmetry,
we work in the simplifying limit $m_q \ll \mu$
and focus on the regime $\mu L \gg 1$.
If $TL \gg 1$, then the sum (\ref{eq:Vfgeneral})
is dominated by the $k = \pm \half$, $n = 1$ terms,
giving
\begin{align}
    V_{\rm f}(\Omega)
    &=
    \frac{\pm2T \, e^{-\nf \pi L T}}{\nf \pi L^2 }
    \left[ \mu \sin(\nf \mu L)+\pi T \cos(\nf \mu L)\right]
\nonumber\\ & {} \times
    \left(\tr \Omega^{\nf } {+}\mathrm{h.c.} \right) + \textrm{(holonomy-independent)},
\label{eq:Vf2}
\end{align}
up to exponentially small corrections.
The $e^{-\pi TL\nf}$ factor
arises from the lowest fermionic Matsubara frequency and
our twisted boundary conditions.
Alternatively, if $TL \to 0$ then
the prefactor in (\ref{eq:Vf2}) becomes
$
    \pm (\nf^2 \pi^2 L^3)^{-1}
$.
In either regime of $TL$, neglecting subdominant contributions,
$V_{\rm eff}(\Omega) \propto \Re \tr \Omega^\nf$ with
an amplitude which oscillates as a function of $\nf \mu L$.

For  $\nf \,{=}\, N \,{=}\, 3$,
extrema of ${V}_{\rm f}(\Omega)$ fall into four categories:
({\it a})
    one center-symmetric extremum at
    $\Omega = \textrm{diag}\,(1,e^{2\pi i /3}, e^{4\pi i /3})$,
    where the $SU(3)$ gauge symmetry is ``broken" down to $U(1)^2$
    with the holonomy playing the role
    of an adjoint Higgs field;
%
({\it b})
    three center-broken extrema with ``residual'' gauge group
    $U(1)^2$ where
    $\Omega = \textrm{diag}\,(e^{(2k-1)i\pi/3},e^{2k i\pi/3},e^{(2k+1)i\pi/3})$,
    $k = 0,1,2$;
({\it c})
    nine center-broken ``$SU(2) \times U(1)$'' extrema at
    $\Omega = \textrm{diag}\,(e^{k i \pi/9},e^{k i \pi/9},e^{-2k i \pi/9})$
    with $k \bmod 6 = 2$, 3 or 4;
({\it d})
    three center-broken ``$SU(3)$'' extrema,
    $\Omega = \textrm{diag}\,(e^{2k i\pi/3},e^{2k i\pi/3},e^{2k i\pi/3})$,
    $k=0,1,2$.
    These ``$SU(3)$'' extrema are also minima of $V_{\rm g}$.

The form (\ref{eq:Vf2})
implies that the 
the locations of the minima of $V_{\rm eff}(\Omega)$ oscillate as a function of
$\mu L$, 
as illustrated in the contour plots in Fig.~\ref{fig:quantum_oscillation}
for two nearby values of $\mu L$.
(Each plot shows a fundamental domain of the Weyl
group of $SU(3)$, which acts by permuting the eigenvalues of $\Omega$.)
There are \emph{quantum oscillations}
in the phase structure of cold dense QCD on a circle,
with the minima of $V_{\rm eff}$ cycling through two inequivalent sets
of local minima as $\mu L$ varies.
These come in two groups within which $V_{\rm f}(\Omega)$ is degenerate.
One group consists of the center-symmetric and three $SU(3)$ extrema.
The other consists of the six $SU(2) \times U(1)$ extrema with
$\Omega = \textrm{diag}\,(e^{k i \pi/9},e^{k i \pi/9},e^{-2k i \pi/9})$
with $k \bmod 6 = 2$ and $4$.
(The remaining six extrema are always saddle points for $TL\gg 1$.)
 
At $T \,{=}\, 0$
there are quantum phase transitions
when the minimum energy state switches from one set of extrema to another,
with associated jumps in the ground state degeneracy.
(Similar behavior in other circle-compactified theories has been seen in
Refs.~\cite{Vshivtsev:1998fg,Kanazawa:2017mgw}.)
There are an infinite number of phase transitions in the cold dense limit
as $L$ increases and successive energy bands pass through the value of
the chemical potential, with an accumulation point at $L = \infty$.
Borrowing a term from the condensed-matter
literature \cite{PhysRevLett.44.1502,Fisher1},
each point in the $(T,\mu)$ phase diagram for QCD
where this phenomenon occurs can be called a multi-phase point
\footnote
    {%
    Unlike the spin models discussed in
    Refs.~\cite{PhysRevLett.44.1502,Fisher1},
    our multi-phase points involve no fine-tuning,
    as they are obtained as limits of one of our
    two symmetry-preserving choices of compactification.
    }.
As we discuss below, this behavior is expected in a finite area domain
of the $(T,\mu)$ phase diagram,
so in fact we find a \emph{multi-phase region}.

The small residual gluon contribution to $V_{\rm eff}$ favors
configurations with clumped holonomy eigenvalues, lowering
the energy of $SU(3)$ extrema relative to the center-symmetric point.
Hence, we expect that all genuine minima of $V_{\rm eff}$
in this multi-phase region
are associated with broken CFC symmetry,
with $\langle \tr \Omega\rangle \neq 0$
\footnote
    {%
    Higher order calculations of $V_{\rm f}$ are needed to confirm
    that the center-broken $SU(3)$ minima are favored,
    as an $\mathcal O(g^2)$ two-loop fermion contribution could
    overwhelm the one-loop gluonic contribution when it is  suppressed
    by Meissner screening.
    }.

Putting everything together, we conclude that there must be some
curve $T = T_{\rm CFC}(\mu)$ below which the CFC symmetry is
spontaneously broken with oscillatory multiphase behavior.
We lack a definitive calculation of $V_{\rm eff}(\Omega)$
valid for $T > T_{\rm CFL}$, but we expect
that $T_{\rm CFC}(\mu)$ is $\mathcal O( g\mu)$,
greatly exceeding $T_{\rm CFL}$ at large $\mu$.
The $T_{\rm CFC}$ curve must end at some point $\mu_c$ on the $T=0$
axis.
The simplest hypothesis is that $\mu_c$ coincides
with $\mu_n \sim \Lambda$, the critical chemical potential needed
to produce pressureless nuclear matter at $T=0$,
as illustrated in Fig.~\ref{fig:phaseDiagram}.

{\bf Conclusions.}  
We have shown that there are well-defined and non-trivial order
parameters for quantum and thermal phase transitions in QCD,
compactified on a circle,
provided $\mathrm{gcd}\,(\nf,N) >1$ with quarks having a common mass $m_q$.
This is a consequence of the existence of color-flavor center symmetry,
and has  interesting implications for the phase structure of QCD
as a function $\nf/N$, $\mu$, $T$, and $m_q$.

There are many worthwhile extensions of our observations.
Consideration of CFC symmetry may be helpful in studies
of QCD behavior near the lower edge of the conformal window
\cite{Kaplan:2009kr}.
For applications to dense QCD, an explicit
calculation of the one-loop gluon contribution to $V_{\rm eff}(\Omega)$
in the hard dense loop approximation would give a better estimate for
the CFC symmetry restoration temperature $T_{\rm CFC} (\mu)$.  The
role of explicit $SU(3)_V$ symmetry breaking should be explored.
Finally, it would be interesting to study local order parameters
for CFC symmetry.  At $\mu = 0$ these order parameters violate
$SU(\nf)_V$ symmetry, and hence must vanish as one takes the
$\mathbb{R}^4$ limit by the Vafa-Witten theorem \cite{Vafa:1983tf}.
But this theorem does not apply at finite $\mu$.

{\bf Acknowledgments.}  We are grateful to D.~B.~Kaplan, L.~McLerran,
S.~Reddy, A.~Roggero, M.~Savage, and T.~Sch\"afer for stimulating discussions.
This work was supported, in part, by the U.~S.~Department of Energy
via grants DE-FG02-00ER-41132 (A.C.), DE-FG02-04ER41338 (S.S.),
DE-SC0013036 (M.\"U.), DE-FG02-00ER41132 (M.W.) and DE-SC0011637
(L.Y.).
L.~Yaffe thanks the University of Regensburg and the Alexander von
Humboldt foundation and S. Sen thanks Los Alamos National Laboratory for their generous support and hospitality during
completion of this work.

\bibliographystyle{apsrev4-1}
\bibliography{small_circle} 

\appendix 

\begin{center} 
{\bf Supplemental Material} 
\end{center} 

{\bf Multiple compactified dimensions.}
Suppose multiple dimensions are compactified, so that the
theory lives on $\mathbb{R}^{D-k}\times T^k$.
If one weakly gauges the $SU(\nf)_V$ flavor symmetry,
then quarks become bifundamentals under $SU(N)\times SU(\nf)$
and there is a $(\mathbb{Z}_{d})^k$ center symmetry
(see, e.g., Refs.~\cite{GonzalezArroyo:2005dz,Kovtun:2007py}).
Charged operators are Wilson loops wrapping non-trivial cycles of $T^{k}$
with non-zero winding numbers mod $d$.
In the limit of vanishing $SU(\nf)$ gauge coupling,
where fluctuations in the $SU(\nf)$ gauge field become negligible,
it is possible to preserve a single diagonal $\mathbb{Z}_d$
subgroup of $(\mathbb{Z}_{d})^k$ by intertwining it with the
$\mathbb{Z}_d$ subgroup of cyclic flavor permutations.
This is achieved by setting
$q(x_{i} {+} L_{i}) = \mathcal{U} \, q(x_i)$,
with $\{x_i\}$ parametrizing $T^k$, and $\mathcal{U}$ given by 
one of the choices (\ref{eq:BCs}).

If $\nf$ and $N$ have multiple common divisors, then one
can choose BCs which preserve different embeddings of
$\mathbb Z_d$ (or a chosen subgroup of $\mathbb{Z}_d$) within $(\mathbb{Z}_N)^k \times SU(\nf)_V$.
As an example, suppose $N = \nf = 4$, with two compactified directions.
Instead of a common boundary condition for both directions, 
one could choose differing flavor-twisted boundary conditions
(\ref{eq:generic_twisted_BCs}) for the two compact directions, with
\begin{align}
    \mathcal{U}_1 \equiv \textrm{diag}(1,-1, 1, -1) \,,\;
    \mathcal{U}_2 \equiv \textrm{diag}(1,1, -1, -1) \,.
\end{align} 
Eigenvalues of these $\mathcal{U}_k$ are transposed under
the action of a $\mathbb{Z}_2 \times \mathbb Z_2$ subgroup
of the $\mathbb Z_4 \times \mathbb Z_4$ center symmetry.
These transpositions can be compensated by 
flavor permutations with a $\mathbb Z_2 \times \mathbb Z_2$
subgroup of the $SU(4)_V$ flavor symmetry,
so these boundary conditions produce a compactified theory
with a $\mathbb{Z}_2 \times \mathbb Z_2$ CFC symmetry in which
each $\mathbb Z_2$ factor affects only a single compact dimension.

\medskip
{\bf Holonomy effective potential on $\mathbb{R}^2\times T^2$.}
Consider a 2-torus $T^2 = S^1_L \times S^1_{\beta}$ with $L$ and
$\beta \equiv 1/T$
regarded as spatial $x_1$ and thermal $x_4$ circle sizes, respectively.
Assign quarks twisted boundary conditions (\ref{eq:BCs}) in the $x_1$
direction, and thermal boundary conditions with a $U(1)_Q$
chemical potential in $x_4$,
\begin{subequations}\label{eq:therm}
\begin{align}
    q(x_4{=}\beta) &= - e^{-\beta \mu} \, q(x_4{=}0) \,,
\\
    \bar q(x_4{=}\beta) &= - e^{\beta \mu} \, \bar q(x_4{=}0) \,.
\end{align}
\end{subequations}
Assume a constant spatial holonomy $\Omega$,
with eigenvalues $\{ e^{i \theta_a} \}$.
In $A_1 = 0$ gauge, the holonomy appears as additional phases in the
$x_1$ boundary condition,%
\begin{subequations}\label{eq:x_1}
\begin{align}
    q(x_1{=}L)_{aA} &=
    e^{i\theta_a} \, \nu^A \, q(x_1{=}0)_{aA} \,,
\\
    \bar q(x_1{=}L)_{aA} &=
    e^{-i\theta_a} \, \nu^{-A} \, \bar q(x_1{=}0)_{aA} \,,
\end{align}
\end{subequations}
where $a = 1,{\cdots},N$ is a color index and
$A$ is a flavor index running from 0 to $\nf{-}1$ for BC (\ref{eq:BCa}),
or a half-integer $\half,{\cdots},\nf{-}\half$ for BC (\ref{eq:BCb}).
The usual quark action,
$
    S_F
    =
    \int d^4x \> \bar{q} \, ( \gamma^\mu \partial_\mu + m_q )\, q
$,
leads to a free energy density
\begin{align}
\label{eq:Ftildedef}
    F
    &=
    - \frac{1}{\beta L \cal V_\perp} \ln \det( \gamma^\mu\partial_\mu + m_q )
\\
    &=
    - \frac{1}{\beta L}
    \int \frac{d^2p_\perp}{(2\pi)^2}
    \sum_{a,n,k}
    \ln \det \big( i \gamma^\mu p^{(a,n,k)}_\mu(\Omega) + m_q \big) ,
\nonumber
\end{align}
where
$
    p_\mu^{(a,n,k)}(\Omega)
    \equiv
    (p_1^{(a,n)}(\Omega),\, \vec p_\perp,\, p_4^{(k)})
$
are momenta consistent with the above boundary conditions.
The thermal (KMS) conditions (\ref{eq:therm}) imply
$
    p_4^{(k)} = 2\pi k T + i\mu
$
with $k \in \mathbb Z{+}\half$.
The flavor twisted conditions (\ref{eq:x_1}) imply
$
    p_1^{(a,n)} = \theta_a/L + 2\pi n / (\nf L) 
$
with $n \in \mathbb Z$ for BC (\ref{eq:BCa}),
or $n \in \mathbb Z{+}\half$ for BC (\ref{eq:BCb}).
Performing the Dirac determinant reduces the free energy density to
\begin{align}
    F 
    &=
    - \frac{2}{\beta L} \int \frac{d^2p_\perp}{(2\pi)^2}
    \sum_{a,n,k}  \ln \!\left( p_\mu^2 +  m_q^2 \right),
\label{eq:Ftilde2}
\end{align}
where the indices labeling the quantized momentum components are suppressed.

To obtain a UV-safe quantity and focus on the effects of the holonomy,
we subtract the holonomy-independent infinite volume limit,
\begin{align}
\label{Fdef}
    \Delta F(\Omega,L)
    &=
    F(\Omega,L)
    - F(\Omega,L\rightarrow\infty)
\nonumber
\\
    &=
    \frac 2\beta
    \sum_k
    \int_0^\infty \frac {dz}{z} \>
    \int \frac{d^2p_\perp}{(2\pi)^2}
\nonumber
\\ &\qquad {} \times
    \left( \frac{1}{L} \sum_{a,n} - \int \frac{dp_1}{2\pi}\right)
    e^{-z(p_\mu^2 + m_q^2)} \,.
\end{align}
The sum-integral difference can be evaluated using Poisson summation,
yielding  
\begin{align}
    \Delta F
    &=
    \frac{1}{4 \pi^{3/2} \beta}
    \int_0^\infty \frac{dz}{z^{5/2}}\> e^{-z m_q^2}
    \sum_{k\in \mathbb Z+\frac 12}
    e^{-z \, (2\pi k/\beta + i\mu)^2}
\nonumber
\\ &\qquad\qquad{}\times
    \sum_{n\ne 0} \> e^{- n^2 L^2/(4z)}
    \sum_{a,A}
    e^{i (\theta_a + 2\pi A/\nf )n}
\nonumber
\\ &=
    \frac{1}{\pi L^3 \beta}
    \sum_{n=1}^\infty \>
    \frac 1{n^3}
    \left(\tr \mathcal{U}^n \, \tr \Omega^{n} + \mathrm{h.c.} \right)
\nonumber
\\ &\qquad\qquad\qquad{}\times
    \sum_{k\in \mathbb{Z}+\frac 12}
	(1 + n L \mathfrak{m}_k) \,
	e^{-n L  \mathfrak{m}_k },
\label{Ffinal}
\end{align}
where the effective mass $\mathfrak m_k$ of each fermion mode is given by
\begin{equation}
    \mathfrak{m}^2_k 
    \equiv
    ( 2\pi k T + i \mu )^2 + m_q^2 \,.
\label{meff}
\end{equation}
The result (\ref{Ffinal}) is the holonomy dependent part of the
fermion contribution to the effective potential $V_{\rm eff}(\Omega)$.
The form (\ref{eq:Vfgeneral}) in the main text follows from evaluating
the traces of the flavor holonomy $\mathcal U$.
At leading order in $m_q/\mu$, one may perform the sum over $k$ and obtain
\begin{align}
    \Delta F
    &=
    \frac{T}{\nf^2 \pi L^3}
    \sum_{n=1}^\infty 
    \frac{(\pm 1)^{n}}{n^3} \,
    \frac{\tr\!\left( \Omega^{\nf n} + \mathrm{h.c.}\right)}
	{\sinh(\pi \nf n \, TL)}
\nonumber
\\ & \quad{} \times
    \Bigl[
	\pi\nf n \, TL \cos(\nf n \,\mu L) \, \coth(\pi \nf n \, TL)
\nonumber
\\ & \qquad {} +
	\nf n \, \mu L \, \sin(\nf n \,\mu L) + \cos( \nf n \, \mu L)
    \Bigr] .
\label{eq:DeltaF}
\end{align}

\medskip
{\bf Zero temperature limit.}
When $TL \to 0$,
the holonomy-dependent free energy (\ref{eq:DeltaF}) reduces to
\begin{align}
   \Delta F
   &=
   \frac{1}{\nf^3 \pi^2 L^4} \sum_{n=1}^\infty
   \frac{(\pm 1)^{n}}{n^4} \,
   \tr \!\left( \Omega^{\nf n} + \mathrm{h.c.}\right)
\nonumber
\\ &\qquad {} \times
     \big[ \nf n \, \mu L \sin(\nf n \mu L) + 2\cos(\nf n \mu L) \big] \,.
\end{align}
In this $TL=0$ limit, both the shape and sign of the
potential depends on $\mu L$.
However, the subsequent terms in the sum over $n$ rapidly
decrease as $1/n^4$.
The extrema of the potential at $TL=0$ are
in the same locations as the extrema at $TL \gg 1$.

\begin{figure*}[p]
\centering
\includegraphics[width=.85\textwidth]{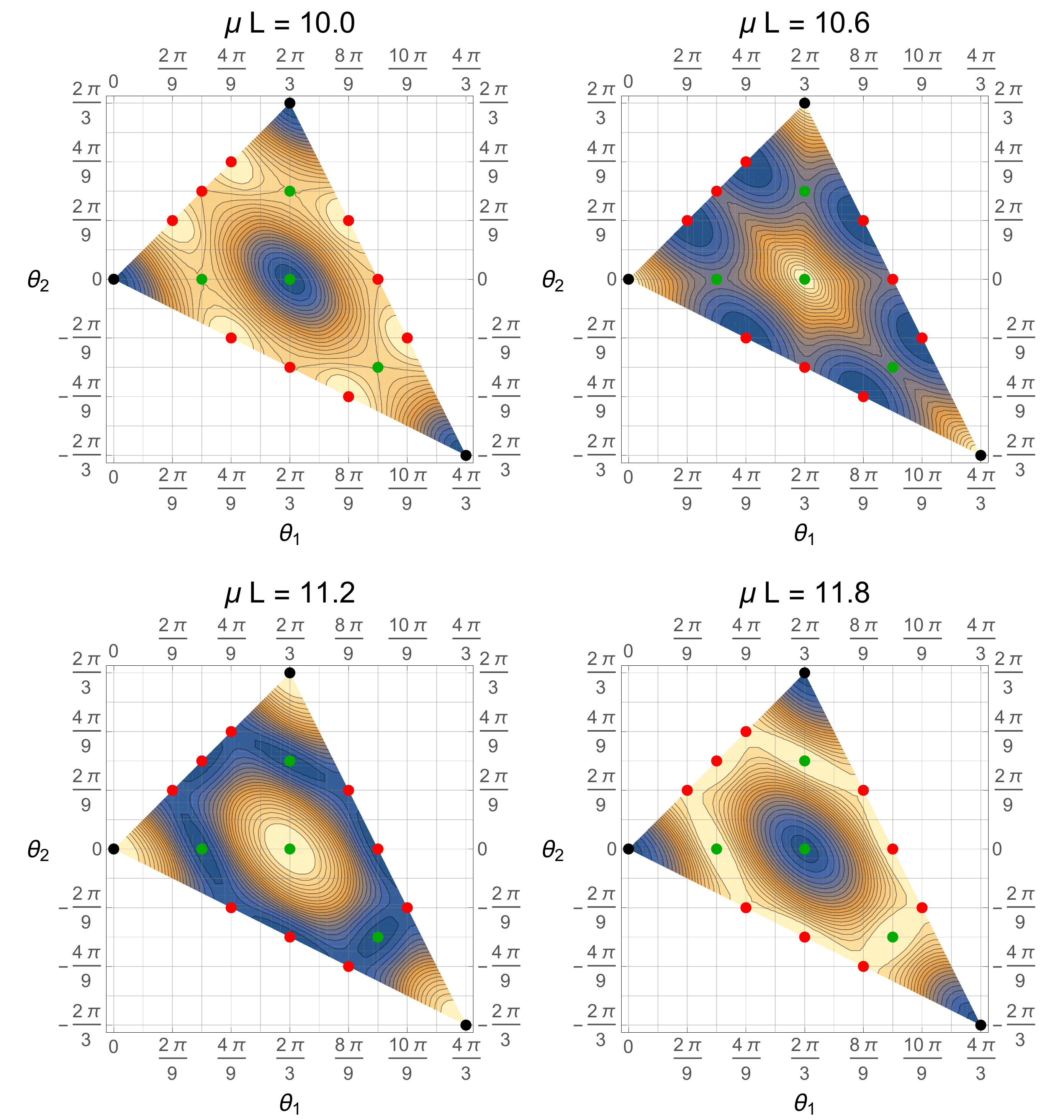}
\vspace*{-10pt}
\caption
    {(Color online.)
    Contour plots of $L^4 V_{\rm eff}(\Omega)$
    as a function of $\theta_1$ and $\theta_2$,
    for $N = \nf = 3$ and $TL = 0$, with BCs \eqref{eq:BCa}.
    Shown are four nearby values of $\mu L$, illustrating the
    existence of quantum oscillations as a function of $\mu L$.
    Darker colors indicate lower values of $L^4 V_{\rm eff}$,
    and regions outside the triangle shown are gauge-equivalent
    to points within the triangle.  
    Dots denote critical points of $V_{\rm f}$.
    Results with BCs \eqref{eq:BCb} are similar.
    }
\label{fig:quantum_oscillation_T0}
\end{figure*}

One finds essentially the same quantum oscillations in the minima
as a function of $\mu L$ at $TL=0$ as in our large $TL$
expressions in the main text.
For completeness, snapshots of
the behavior at $TL = 0$,
for $N = \nf = 3$ and several nearby values of $\mu L$,
are shown in Fig.~\ref{fig:quantum_oscillation_T0}.
The figure highlights one qualitative difference between the
$TL = 0$ and $TL \gg 1$ results.
As noted in the main text, for $TL \gg 1$, six of the
center-breaking extrema are always saddlepoints of the
effective potential, regardless of the value of $\mu L$.
But at $TL=0$ these extrema also take turns as minima
of the potential as $\mu L$ varies.
Hence, unsurprisingly, there is non-uniformity between the $T \to 0$
and $L \to \infty$ limits.
If one views the spatial compactification purely as a device used
to probe the behavior of the system, then it is natural to choose
$L$ large compared to all other physical length scales.
This motivated our emphasis on the large $TL$ regime in the main text.

\end{document}